\documentclass{elsart}

\usepackage[english]{babel}
\usepackage{amsmath,amssymb,mathrsfs}
\usepackage[dvips]{graphicx}
\usepackage[square,comma,sort&compress]{natbib}

\newcommand{\ie}{\textit{i.e.}}
\newcommand{\eg}{\textit{e.g.}}

\newcommand{\sq}{\sin^2 2\theta}
\newcommand{\dm}{\Delta m^2}

\newcommand{\Epr}{E_\mathrm{pr}}

\begin{document}

\raisebox{8mm}[0pt][0pt]{\hspace{10.5cm}
\vbox{\tt TUM-HEP-524/03}}

\begin{frontmatter}

\title{Variations on KamLAND: 
  likelihood analysis and frequentist confidence regions}

\author{Thomas Schwetz} 
\ead{schwetz@ph.tum.de}

\address{Institut f{\"u}r Theoretische Physik, Physik Department, 
  Technische Universit{\"a}t M{\"u}nchen,
  James--Franck--Strasse, D--85748 Garching, Germany}

\begin{abstract}
  In this letter the robustness of the first results from the KamLAND
  reactor neutrino experiment with respect to variations in the
  statistical analysis is considered. It is shown that an
  event-by-event based likelihood analysis provides a more powerful
  tool to extract information from the currently available data sample
  than a least-squares method based on energy binned data.
  Furthermore, a frequentist analysis of KamLAND data is performed.
  Confidence regions with correct coverage in the plane of the
  oscillation parameters are calculated by means of a Monte Carlo
  simulation. I find that the results of the usually adopted
  $\chi^2$-cut approximation are in reasonable agreement with the exact
  confidence regions, however, quantitative differences are detected. 
  Finally, although the current data is consistent with an energy
  independent flux suppression, a $\sim 2\sigma$ indication in favour
  of oscillations can be stated, implying quantum mechanical
  interference over distances of the order of 200~km.
\end{abstract}
 

\end{frontmatter}

\section{Introduction}

The outstanding results from the KamLAND reactor neutrino
experiment~\cite{kamlandPRL} have lead to a significant progress in
neutrino physics. The observed disappearance of reactor anti-neutrinos
is in agreement with the so-called LMA solution of the solar neutrino
problem~\cite{sol_ex}. Alternative oscillation solutions like LOW, VAC
or SMA are ruled out with very high confidence
level~\cite{Maltoni:2002aw,Fogli:2002au,Bahcall:2002ij,Bandyopadhyay:2002en,deHolanda:2002iv,Creminelli:2001ij,Aliani:2002na,Barger:2002at,Nunokawa:2002mq,Balantekin:2003dc},
and non-oscillation mechanisms can play only a sub-leading role (for
a review and references see Ref.~\cite{Pakvasa:2003zv}).

These important conclusions are based on a data sample consisting of
54 events above the geo-neutrino threshold in KamLAND. The purpose of
this letter is to discuss issues related to the statistical analysis
of these data. In Sec.~\ref{sec:KL_analyses} an event-by-event based
likelihood analysis is compared to the widely used least-squares
method based on energy binned data. It is shown that the likelihood
method allows one to extract more precise information about the
oscillation parameters from KamLAND data. Since the currently
available data sample consists only of rather few events, one might
ask the question whether the approximate confidence regions obtained
from the usual $\chi^2$-cut method are reliable. In Sec.~\ref{sec:FC}
this question is addressed by calculating frequentist confidence
regions for the oscillation parameters according to the prescription
given by Feldman and Cousins~\cite{FC}.  The explicit
construction of the confidence regions by Monte Carlo simulation takes
properly into account statistical fluctuations of the rather small
data sample and the non-linear character of the oscillation
parameters. In Sec.~\ref{sec:decoherence} the statistical significance
of an oscillatory signal in the KamLAND data is discussed, and I
conclude in Sec.~\ref{sec:conclusions}.

\section{Comparing likelihood and least-squares methods}
\label{sec:KL_analyses}

The current KamLAND data sample consists of 86 anti-neutrino events in the
full energy range. In the lower part of the spectrum there is a relevant
contribution from geo-neutrino events to the signal.  To avoid large
uncertainties associated with the geo-neutrino flux an energy cut at 2.6~MeV
prompt energy is applied for the oscillation analysis, and 54 anti-neutrino
events remain in the final sample. All analyses of KamLAND
data~\cite{Maltoni:2002aw,Fogli:2002au,Bahcall:2002ij,Bandyopadhyay:2002en,deHolanda:2002iv,Creminelli:2001ij,Aliani:2002na,Barger:2002at,Nunokawa:2002mq,Balantekin:2003dc,Fiorentini:2003ww,Ianni:2003xy}
performed so far outside the KamLAND collaboration are using these data binned
into 13 energy intervals above the geo-neutrino cut, as given in Fig.~5 of
Ref.~\cite{kamlandPRL}.\footnote{For an analysis including the geo-neutrino
events see Ref.~\cite{Fiorentini:2003ww}.} In Subsec.~\ref{sec:likelihood} I
describe an alternative analysis based on the likelihood function of the data,
which allows one to take into account the precise energy information contained
in each single event. The results of this analysis are compared to the ones
from the energy binned least-squares method in Subsec.~\ref{sec:chisq}.

Before exact confidence regions are calculated in Sec.~\ref{sec:FC}
the usual $\chi^2$-cut approximation will be used. One constructs a
statistic $\Delta X^2(\sq, \dm)$ from the data. Under certain
assumptions, like large sample limit and linear
parameter dependence, this statistic will be distributed as a $\chi^2$
with 2 degrees of freedom, independent of the point in the parameter
space~\cite{books,pdg}.  Then a given point $(\sq, \dm)$ is contained
in the allowed region at $\beta$ CL if
\begin{equation}\label{eq:Dchi2}
\Delta X^2(\sq,\dm) \le \Delta\chi^2_\beta(2) 
\,, \quad\mbox{where}\quad
\int_0^{\Delta\chi^2_\beta(n)} f_{\chi^2} ( x, n ) {\rm d} x = \beta \,.
\end{equation}
Here $f_{\chi^2} ( x, n )$ denotes the $\chi^2$-distribution with $n$
degrees of freedom.  In the following I will refer to this procedure as
``$\chi^2$-cut method''. In this section it will be applied to
calculate approximate confidence regions by using the likelihood as
well as the least-squares method.

\subsection{Likelihood analysis of KamLAND data}
\label{sec:likelihood}

For given oscillation parameters $\sq$ and $\dm$ the predicted event
spectrum in KamLAND can be calculated by
\begin{align}\label{eq:spect}
    f(\Epr; & \, \sq,\dm) = \nonumber \\
    & \mathcal{N} 
    \int_0^\infty \mathrm{d}E_\nu  \, \sigma(E_\nu) 
    \sum_j \phi_j(E_\nu)  P_j(E_\nu, \sq, \dm) 
    \, R(\Epr,\Epr')\,.
\end{align}
Here $R(\Epr, \Epr')$ is the energy resolution function and $\Epr,
\Epr'$ are the observed and the true prompt energies, respectively,
and we use a Gaussian energy resolution of
$7.5\%/\sqrt{\Epr(\mathrm{MeV})}$~\cite{kamlandPRL}. The neutrino
energy is related to the true prompt energy by $E_\nu = \Epr' + \Delta
- m_e$, where $\Delta$ is the neutron-proton mass difference and $m_e$
is the positron mass.  The cross section $\sigma(E_\nu)$ for the
detection process $\bar\nu_e + p \to e^+ + n$ is taken from
Ref.~\cite{Vogel:1999zy}. The neutrino spectrum $\phi(E_\nu)$ from
nuclear reactors is well known. I am using the phenomenological
parameterisation from Ref.~\cite{reactor_spect} and the average fuel
composition for the nuclear reactors as given in
Ref.~\cite{kamlandPRL}. The sum over $j$ in Eq.~\eqref{eq:spect} runs
over 16 nuclear plants, taking into account the different distances
from the detector $L_j$ and the power output of each reactor (see
Table~3 of Ref.~\cite{kamlandproposal}).  Finally, $P_j(E_\nu, \sq,
\dm)$ is the survival probability for neutrinos emitted at the reactor
$j$, depending on the distance $L_j$, the neutrino energy and the
two-flavour oscillation parameters $\sq$ and $\dm$.

The total number of events predicted for oscillation parameters $\sq$
and $\dm$ above the geo-neutrino cut $E_\mathrm{cut} = 2.6$~MeV is given by
\begin{equation}
N_\mathrm{pred}(\sq,\dm) = \int_{E_\mathrm{cut}}^\infty 
\mathrm{d} \Epr \, f(\Epr; \, \sq , \dm) \,.
\end{equation}
The over-all constant $\mathcal{N}$ in Eq.~(\ref{eq:spect}) is
determined by normalising the number of events for no oscillations to
$N_\mathrm{pred}(\sq = 0,\dm = 0) = 86.8$~\cite{kamlandPRL}. The
probability distribution of the expected events, \ie\ the probability
to obtain an event with the prompt energy $\Epr$ in the interval
$[\Epr, \Epr + \mathrm{d} \Epr]$, can be obtained by normalising the
spectrum given in Eq.~(\ref{eq:spect}):
\begin{equation}\label{eq:pdf}
p(\Epr; \, \sq,\dm) = 
\frac{f(\Epr; \, \sq,\dm)}{N_\mathrm{pred}(\sq,\dm)}\,.
\end{equation}

\begin{table}
\begin{tabular}{|@{\quad}r@{\quad}c@{\quad}|@{\quad}r@{\quad}c@{\quad}|@{\quad}r@{\quad}c@{\quad}|@{\quad}r@{\quad}c@{\quad}|}
\hline
$i$ & $\Epr^i$ [MeV] &
$i$ & $\Epr^i$ [MeV] &
$i$ & $\Epr^i$ [MeV] &
$i$ & $\Epr^i$ [MeV] \\
\hline
1  & 0.906 & 23 &   2.151 &45 &   3.243 & 67 &   4.284 \\[-3mm]
2  & 0.978 & 24	&   2.280 &46 &   3.328 & 68 &   4.322 \\[-3mm]
3  & 1.035 & 25	&   2.294 &47 &   3.345 & 69 &   4.353 \\[-3mm]
4  & 1.089 & 26	&   2.314 &48 &   3.382 & 70 &   4.414 \\[-3mm]
5  & 1.198 & 27	&   2.524 &49 &   3.416 & 71 &   4.420 \\[-3mm]
6  & 1.205 & 28	&   2.531 &50 &   3.437 & 72 &   4.455 \\[-3mm]
7  & 1.208 & 29	&   2.534 &51 &   3.460 & 73 &   4.577 \\[-3mm]
8  & 1.262 & 30	&   2.565 &52 &   3.484 & 74 &   4.610 \\[-3mm]
9  & 1.313 & 31	&   2.568 &53 &   3.504 & 75 &   4.675 \\[-3mm]
10 & 1.340 & 32	&   2.595 &54 &   3.650 & 76 &   4.726 \\[-3mm]
11 & 1.378 & 33	&   2.636 &55 &   3.671 & 77 &   4.804 \\[-3mm]
12 & 1.408 & 34	&   2.721 &56 &   3.671 & 78 &   4.997 \\[-3mm]
13 & 1.524 & 35	&   2.782 &57 &   3.718 & 79 &   5.021 \\[-3mm]
14 & 1.639 & 36	&   2.843 &58 &   3.735 & 80 &   5.150 \\[-3mm]
15 & 1.683 & 37	&   2.850 &59 &   3.864 & 81 &   5.160 \\[-3mm]
16 & 1.703 & 38	&   2.982 &60 &   3.881 & 82 &   5.269 \\[-3mm]
17 & 1.748 & 39	&   2.982 &61 &   3.915 & 83 &   5.289 \\[-3mm]
18 & 1.812 & 40 &   3.040 &62 &   3.969 & 84 &   5.482 \\[-3mm]
19 & 1.832 & 41	&   3.060 &63 &   4.115 & 85 &   5.689 \\[-3mm]
20 & 1.985 & 41	&   3.162 &64 &   4.142 & 86 &   5.706 \\[-3mm]
21 & 2.029 & 43	&   3.226 &65 &   4.261 &    &         \\[-3mm]
22 & 2.100 & 44 &   3.240 &66 &   4.268 &    &         \\
\hline
\end{tabular}
\caption{Prompt energies of the 86 anti-neutrino events in KamLAND.}
\label{tab:events} 
\end{table}

The prompt energies of the 86 events observed in KamLAND can be
extracted from Fig.~3 of Ref.~\cite{kamlandPRL} and are listed in
Tab.~\ref{tab:events}. Using the 54 events above the geo-neutrino cut
with $\Epr^i > E_\mathrm{cut}$ one obtains the likelihood function
containing the spectral shape information of the data:
\begin{equation}
\mathcal{L}_\mathrm{shape}(\sq, \dm) = 
\prod_{i=33}^{86} p(\Epr^i; \, \sq,\dm) \,.
\end{equation}
To take into account also the information implied by the
total number of observed events I apply the modified likelihood
method (see, \eg, Ref.~\cite{books}):
\begin{equation}\label{eq:Ltot}
\mathcal{L}_\mathrm{tot}(\sq, \dm) = 
\mathcal{L}_\mathrm{shape}(\sq, \dm) \times
\mathcal{L}_\mathrm{rate}(\sq, \dm)   
\end{equation}
with\footnote{In general a Poisson distribution has to be used for
  $\mathcal{L}_\mathrm{rate}$. However, for a mean of order 
  $N_\mathrm{obs} = 54$ the Poisson distribution is very well
  approximated by the Gaussian distribution.}
\begin{equation}
\mathcal{L}_\mathrm{rate}(\sq, \dm) = 
\frac{1}{\sqrt{2\pi} \sigma_\mathrm{rate}}
\exp\left[ -\frac{1}{2} \left(
\frac{N_\mathrm{pred}(\sq,\dm) - N_\mathrm{obs}}{\sigma_\mathrm{rate}}
\right)^2 \right] \,.
\end{equation}
Here $N_\mathrm{obs} = 54$ is the observed number of events, and 
\begin{equation}\label{eq:sigma}
\sigma_\mathrm{rate}^2 (\sq,\dm) = 
N_\mathrm{pred}(\sq,\dm) + \sigma_\mathrm{syst}^2
N_\mathrm{pred}^2(\sq,\dm) \,, 
\end{equation}
with the systematical error $\sigma_\mathrm{syst} = 6.42\%$~\cite{kamlandPRL}.
Note that we derive $\sigma_\mathrm{rate}$ from the predicted number
of events, which introduces the parameter dependence of
$\sigma_\mathrm{rate}$. 

\begin{figure}
\begin{center} 
    \includegraphics*[width=0.7\linewidth]{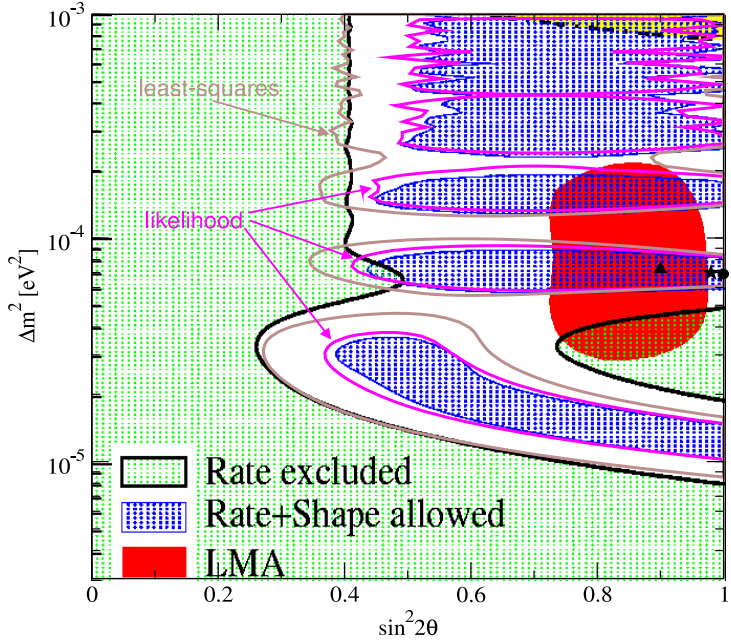}
\end{center}
\vspace*{-1.5cm}
    \caption{ Comparison of the 95\% CL regions obtained from the
        likelihood and the least-squares methods (lines) with the ones
        published by the KamLAND collaboration~\cite{kamlandPRL}
        (regions labeled as ``Rate+Shape allowed''). The best fit
        points are marked by a star, triangle, dot for the likelihood,
        least-squares, KamLAND analyses, respectively.}
  \label{fig:overlay}
\end{figure}

By maximising the likelihood function Eq.~(\ref{eq:Ltot}) the best fit
parameters $\dm = 7.05 \times 10^{-5}$eV$^2$ and $\sq = 0.98$ are
obtained, in very good agreement with the values obtained by the
KamLAND collaboration: $\dm = 6.9 \times 10^{-5}$eV$^2$ and $\sq =
1$~\cite{kamlandPRL}. To calculate allowed regions for the parameters
by means of the $\chi^2$-cut method one defines~\cite{books,pdg}
\begin{equation}\label{eq:dXLH}
\Delta X^2(\sq, \dm) 
= 2 \ln\mathcal{L}_\mathrm{tot,max} 
- 2 \ln\mathcal{L}_\mathrm{tot}(\sq,\dm) \,, 
\end{equation}
where $\mathcal{L}_\mathrm{tot,max}$ is the maximum of the likelihood
function with respect to $\sq$ and $\dm$. The 95\% confidence regions
obtained from Eq.~(\ref{eq:dXLH}) according to Eq.~(\ref{eq:Dchi2})
are shown in Fig.~\ref{fig:overlay}. One finds that they are in
excellent agreement with the ones published by the KamLAND
collaboration.

\subsection{Least-squares analysis of KamLAND data}
\label{sec:chisq}

The least-squares analyses performed by many
authors~\cite{Maltoni:2002aw,Fogli:2002au,Bahcall:2002ij,Bandyopadhyay:2002en,deHolanda:2002iv,Creminelli:2001ij,Aliani:2002na,Barger:2002at,Nunokawa:2002mq,Balantekin:2003dc}
are based on the KamLAND data binned into 13 energy intervals,
$N^i_\mathrm{obs},\,i=1,\ldots,13$, as given in Fig.~5 of
Ref.~\cite{kamlandPRL}.\footnote{In Ref.~\cite{Ianni:2003xy} a different
likelihood analysis of KamLAND data has been presented, including a goodness
of fit evaluation by Monte Carlo methods. Note however, that in
Ref.~\cite{Ianni:2003xy} the likelihood function is also calculated from the
energy binned data similar to the least-squares method, in contrast to the
event-by-event based likelihood discussed in the present work.} Since the
number of events in the individual bins is rather small (in some bins even
zero) the use of a least-squares statistic based on the Poisson distribution
is appropriate~\cite{pdg}:
\begin{equation}\label{eq:pois}
X^2(\sq , \dm) = 2 \sum_i \left\{
\alpha N^i_\mathrm{pred} - N^i_\mathrm{obs} +
N^i_\mathrm{obs} \ln  
\frac{N^i_\mathrm{obs}}{\alpha N^i_\mathrm{pred}}
\right\} +
\left( \frac{1-\alpha}{\sigma_\mathrm{syst}} \right)^2 \,,
\end{equation}
where the term containing the logarithm is absent in bins with no
events. The predicted number of events $N^i_\mathrm{pred}(\sq,\dm)$ in
bin $i$ is obtained by integrating the spectrum Eq.~(\ref{eq:spect})
over the prompt energy interval corresponding to that bin.
Eq.~(\ref{eq:pois}) has to be minimised with respect to $\alpha$ in
order to take into account the overall uncertainty of the theoretical
predictions $\sigma_\mathrm{syst} = 6.42\%$~\cite{kamlandPRL}.
Although the ``least-squares'' character of the statistic $X^2$ in
Eq.~(\ref{eq:pois}) is not explicitly visible due to the use of the
Poisson distribution it is denoted here by this term to stress the
analogy to the commonly used ``$\chi^2$-function''. A comparison of
KamLAND analyses using Gaussian and Poisson least-squares functions can
be found in Ref.~\cite{Maltoni:2002aw}.

The best fit parameters obtained by minimising Eq.~(\ref{eq:pois}) are
$\dm = 7.24 \times 10^{-5}$~eV$^2$ and $\sq = 0.90$.  Assuming that
\begin{equation}
\Delta X^2(\sq,\dm) = X^2(\sq,\dm) - X^2_\mathrm{min}
\end{equation}
follows a $\chi^2$-distribution with 2 degrees of freedom approximate
confidence regions are obtained by considering contours of constant $\Delta
X^2$ according to the $\chi^2$-cut method in Eq.~(\ref{eq:Dchi2}). From
Fig.~\ref{fig:overlay} we find that the 95\% confidence regions obtained by
this method are significantly larger than the ones from the likelihood
analysis and the regions published by the KamLAND collaboration. This fact was
already noted in Ref.~\cite{Fogli:2002au}.  I conclude that the loss of
information implied by the binning of the data is not negligible, and the
likelihood analysis provides a more powerful method to extract information
from the current KamLAND data sample. Note, however, that in future the
differences between the two methods are expected to decrease, since if more
data is available a smaller bin size can be chosen, and in the limit of zero
bin width the least-squares method converges to the un-binned likelihood
method.

\section{Confidence regions with correct coverage}
\label{sec:FC}

Since the number of events in the currently available KamLAND data sample is
rather small the question arises, whether the standard procedures to calculate
confidence regions as described in Sec.~\ref{sec:KL_analyses} are reliable.
Especially the assumption concerning the distribution of $\Delta X^2$ might be
not justified. Moreover, the parameters of interest, $\dm$ and $\sq$, enter
the problem in a highly non-linear way, which leads to multiple local maxima
of the likelihood function (or local minima of the $X^2$-statistic). In such a
case the actual confidence level of the parameter regions obtained from
Eq.~(\ref{eq:Dchi2}) may differ significantly from the canonical value
$\beta$. To check the robustness of the results I have calculated frequentist
confidence regions, where the correct coverage is guaranteed by construction.
To this aim I follow the prescription given by Feldman and Cousins in
Ref.~\cite{FC}.

\begin{figure}
    \includegraphics[width=0.98\linewidth]{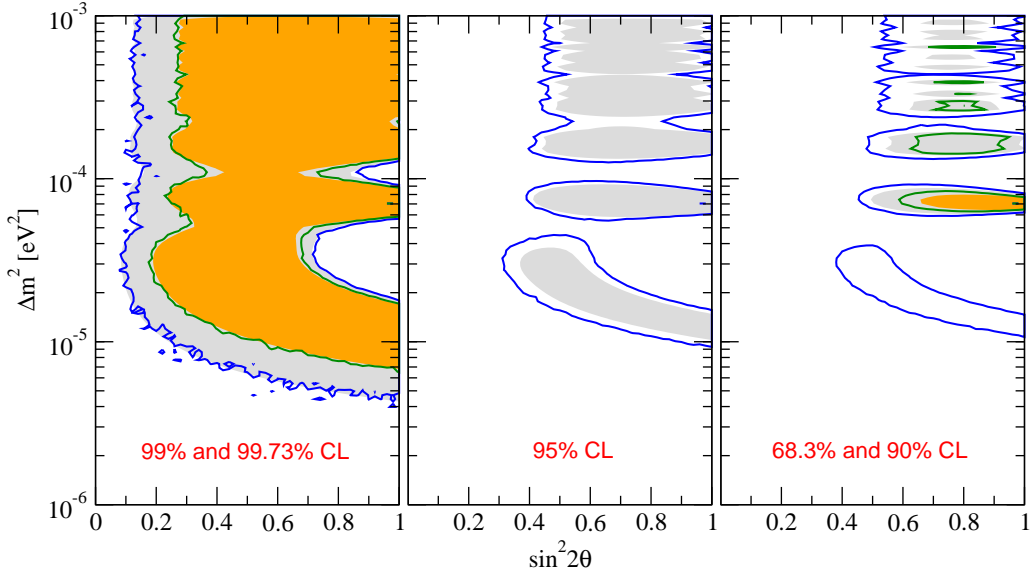}
    \caption{\label{fig:fc_regions_lh}%
        Comparison of confidence regions with correct coverage (lines)
        with the regions obtained from the $\chi^2$-cut approximation
        (shaded regions) for the likelihood method.}
\end{figure}

For both approaches discussed above -- likelihood as well as least-squares
methods -- many synthetic data sets are simulated for fixed
oscillation parameters. 
In the case of the likelihood method first the number of events in the
synthetic data sample $N_\mathrm{sim}(\sq,\dm)$ is generated from a
Gaussian distribution with mean $N_\mathrm{pred}(\sq,\dm)$ and
standard deviation $\sigma_\mathrm{rate}$ given in 
Eq.~(\ref{eq:sigma}). Then the prompt energies of the $N_\mathrm{sim}$
events is thrown according to the distribution Eq.~(\ref{eq:pdf}).
To test the least-squares method a value $\alpha_\mathrm{sim}$
for the parameter $\alpha$ describing the normalisation uncertainty in
Eq.~(\ref{eq:pois}) is generated from a Gaussian distribution with mean 1 and
standard deviation $\sigma_\mathrm{syst}$. Then the number of events
in each bin $i$ is simulated from a Poisson distribution with the mean
$\alpha_\mathrm{sim} N_\mathrm{pred}^i(\sq,\dm)$.

Each ``data set'' generated this way is analysed as described in
Sec.~\ref{sec:KL_analyses} in order to calculate $\Delta
X^2(\sq,\dm)$. For each point on a sufficiently dense grid in the
$(\sq,\dm)$ plane this has been done $10^4$ times for the likelihood
method and $10^5$ times for the least-squares method to map out the actual
distribution of $\Delta X^2$ in that point: $p_\mathrm{sim}(\Delta
X^2;\, \sq,\dm)$. Then, in analogy to Eq.~(\ref{eq:Dchi2}), the point
$(\sq,\dm)$ is included in the confidence region at $\beta$ CL if
$\Delta X_\mathrm{data}^2(\sq,\dm)$ obtained from the real data is
smaller than the one of $100\beta\%$ of the simulated data sets in
that point in the parameter space:
\begin{equation}\label{eq:Xqbeta}
\Delta X_\mathrm{data}^2(\sq,\dm) \le X^2_\beta\,, \quad\mbox{where}\quad
\int_0^{X^2_\beta} p_\mathrm{sim}(x;\, \sq,\dm ) {\rm d} x = \beta \,.
\end{equation}

\begin{figure} 
    \includegraphics[width=0.98\linewidth]{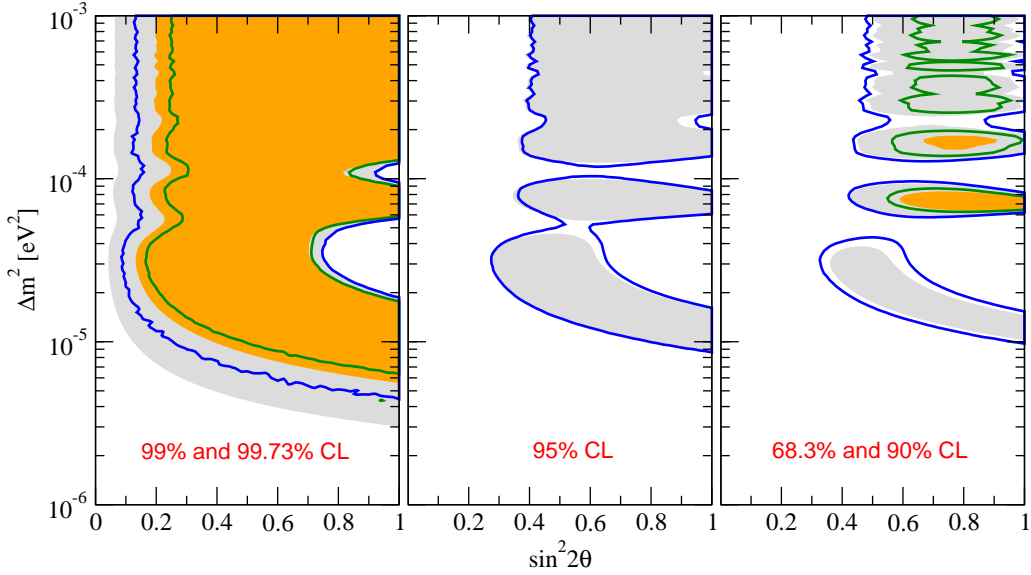}
    \caption{\label{fig:fc_regions_chi2}%
        Comparison of confidence regions with correct coverage (lines)
        with the regions obtained from the $\chi^2$-cut approximation
        (shaded regions) for the least-squares method.}
\end{figure}

The results of these analyses are shown in Figs.~\ref{fig:fc_regions_lh} and
\ref{fig:fc_regions_chi2} for the likelihood and the least-squares methods,
respectively. The regions with correct coverage are compared to the ones
obtained by the $\chi^2$-cut approximation. In both cases reasonable agreement
of the exact and approximate confidence regions is found, although
quantitative differences are visible. 
For the likelihood method the regions at 99.73\% and 99\% CL are in
excellent agreement, whereas for lower confidence levels the
$\chi^2$-cut approximation gives regions somewhat smaller than the
exact ones. Especially the region $10^{-5}$~eV$^2 \lesssim \dm
\lesssim 4\times 10^{-5}$~eV$^2$ does not appear at 90\% CL for the
$\chi^2$-cut approximation.
In the case of the least-squares method the 95\% CL regions are in
excellent agreement. For higher confidence levels the $\chi^2$-cut
approximation gives regions somewhat larger than the exact ones,
whereas for lower confidence levels the allowed regions are a bit
underestimated. {\it E.g.}, the region $\dm \gtrsim 2\times
10^{-4}$~eV$^2$ does not appear at 68.3\% CL for the $\chi^2$-cut
approximation.
In general the $\chi^2$-cut approximation works quite well in the vicinity
of the best fit point around $\dm \approx 7\times 10^{-5}$~eV$^2$.

\begin{figure} 
    \begin{center}
      \includegraphics[width=0.7\linewidth]{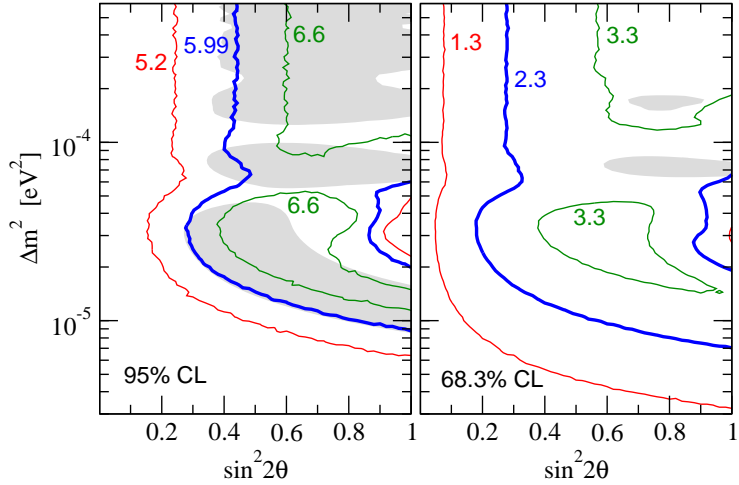}
    \end{center}
    \caption{\label{fig:delta_fc}%
        Contours of constant $X^2_\beta$ as defined in Eq.~(\ref{eq:Xqbeta})
	for the least-squares method for $\beta = 0.95$ (left panel) and
	$\beta = 0.683$ (right panel). The thick lines correspond to the
	values which would result from a $\chi^2$-distribution with 2 degrees
	of freedom. The shaded regions are the approximate allowed regions
	from the $\chi^2$-cut method.}
\end{figure}

A better understanding of these results can be obtained by considering how
$X^2_\beta$ defined in Eq.~(\ref{eq:Xqbeta}) varies as a function of the
oscillation parameters. Note that this quantity does {\it not} depend on the
actually observed data; it characterises the properties of the statistical
method applied to the specific experimental setup. For definiteness I consider
the least-squares method,\footnote{Similar behaviour is also found for the
likelihood method.} for which contours of constant $X^2_\beta$ are shown in
Fig.~\ref{fig:delta_fc} for $68.3\%$ and $95\%$~CL. In the left panel of that
figure one can see that the contour for $X^2_{0.95} = 5.99$, which corresponds
to the $\chi^2$-distribution for 95\% CL, happens to be rather close to the
95\% CL region from the $\chi^2$-cut approximation. This explains the good
agreement observed in the middle panel of Fig.~\ref{fig:fc_regions_chi2}.
Furthermore, one finds from Fig.~\ref{fig:delta_fc} that $X^2_\beta$ decreases
for small values of $\sin^22\theta$ and $\Delta m^2$. The reason for this
behaviour can be understood as follows: If $\sin^22\theta$ becomes $\lesssim
0.2$ and/or $\Delta m^2 \lesssim 10^{-5}$~eV$^2$ the effect of oscillations
gets very small, and the signal in KamLAND corresponds roughly to the
no-oscillation case. Analysing data generated from parameters in that region
leads to best fit points also in the no-oscillation region, with a rather
similar $X^2$. Hence, the distribution of $\Delta X^2$ is more peaked at low
values, which implies relatively small values of $X^2_\beta$.  This explains
the tendency of more constraining exact confidence regions for small values of
$\sin^22\theta$ and $\Delta m^2$. The physical reason for this behaviour is
that even with the present KamLAND data sample no-disappearance can be very
well distinguished from oscillations with $\sin^22\theta \gtrsim 0.2$ and
$\Delta m^2 \gtrsim 10^{-5}$~eV$^2$. Moreover, in the region of small
$\sin^22\theta$ or $\Delta m^2$ the full 2-parameter dependence of the survival
probability is lost, and the effective number of degrees of freedom is
reduced, leading to smaller values of $X^2_\beta$.
The reason why for the likelihood method the agreement is better for higher CL
than for the least-squares method can be partially attributed to the fact that
for the latter the approximate regions extend to smaller values of
$\sin^22\theta$, which implies smaller values of $X^2_\beta$ and larger
disagreement with the $\chi^2$-approximation.

In contrast, for $\sin^22\theta \gtrsim 0.4$ and $\Delta m^2 \gtrsim
10^{-5}$~eV$^2$ the simulation yields relatively higher values of $X^2_\beta$.
In that region oscillations are important. However, because of the rather
small data sample the signature of given parameters can not be identified with
sufficiently high significance. This implies that statistical fluctuations can
lead easily to best fit points in a different local minimum.  In other words,
when data are generated by given parameters in that region, fluctuations can
mimic a signal which is better fitted by quite different parameters. Hence,
the best fit points are stronger affected by fluctuations, resulting into a
broader distribution of $\Delta X^2$ and larger values of $X^2_\beta$. From
Fig.~\ref{fig:delta_fc} one observes that the approximate confidence regions
at 68.3\% CL are well inside the high $X^2_\beta$ regime. This explains the
weaker constraints from exact allowed regions for low confidence levels in
Fig.~\ref{fig:fc_regions_chi2}.
Note, however, that in the region around $\Delta m^2 \sim 7\times
10^{-5}$~eV$^2$, where KamLAND is most sensitive to oscillations, the
$X^2_\beta$ values decrease again. This shows that in that region KamLAND can
well identify the parameters, leading already to statistical properties close
to the expected $\chi^2$-distribution. Therefore, around the best fit point
exact and approximate confidence regions are in good agreement.

The main conclusion to be drawn from Figs.~\ref{fig:fc_regions_lh} and 
\ref{fig:fc_regions_chi2} is that already using the current 54 events from
KamLAND the approximate $\chi^2$-cut method gives a rather reliable
determination of allowed regions. The small quantitative differences to exact
confidence regions can be understood by considering the impact of statistical
fluctuations on the fit. One may expect that once more data will have been
collected the differences will further decrease, because statistical
fluctuations will be less important. Furthermore, if the true oscillation
parameters happen to be close to the present best fit region around $\Delta
m^2 \sim 7\times 10^{-5}$~eV$^2$ and large $\sin^22\theta$ a rather clear
oscillation signal can be observed in KamLAND. In that case the impact of
other local minima will become small and one may expect that $\Delta X^2$ will
follow a distribution rather close to the $\chi^2$-distribution.

\section{Have we already observed oscillations in KamLAND?}
\label{sec:decoherence}

Because of the limited statistics of the current KamLAND data sample
the data is consistent with an energy independent suppression of the
reactor neutrino flux~\cite{kamlandPRL}. This is evident, since
allowed regions appear for large $\dm$, corresponding to energy
averaged oscillations. However, even this first data set indicates
some spectral distortion which is consistent with neutrino
oscillations. In this section the statistical significance for an
oscillatory signal is quantified by using a so-called decoherence
parameter. The survival probability for the electron anti-neutrinos is
modified (in a rather arbitrary way) by multiplying the quantum
mechanical interference term which leads to the oscillatory behaviour
by a factor $(1-\zeta)$:
\begin{equation}\label{eq:decoh}
  P = 1 - \frac{1}{2} \, \sq 
  \left[ 1 - ( 1 - \zeta ) \cos \frac{\dm L}{2 E_\nu} \right] \,. 
\end{equation}
Restricting $\zeta$ to the interval $[0,1]$ one can describe in a
model independent way a loss of quantum mechanical coherence due to
some unspecified mechanism. $\zeta = 0$ corresponds to standard
quantum mechanics, whereas $\zeta = 1$ describes complete decoherence,
\ie, an energy and baseline independent suppression of the
flux.\footnote{Decoherence might have its origin {\it e.g.}\ in quantum
gravity~\cite{Ellis:1983jz}. See also Ref.~\cite{lisi} and references
therein, for an application to atmospheric neutrino oscillations. A
method similar to Eq.~(\ref{eq:decoh}) has been used in
Ref.~\cite{decoh} to investigate the evidence for quantum mechanical
interference in the $B^0 \overline B^0$ and $K^0 \overline K^0$
systems.}

\begin{figure}
\begin{center} 
  \includegraphics*[width=0.65\linewidth]{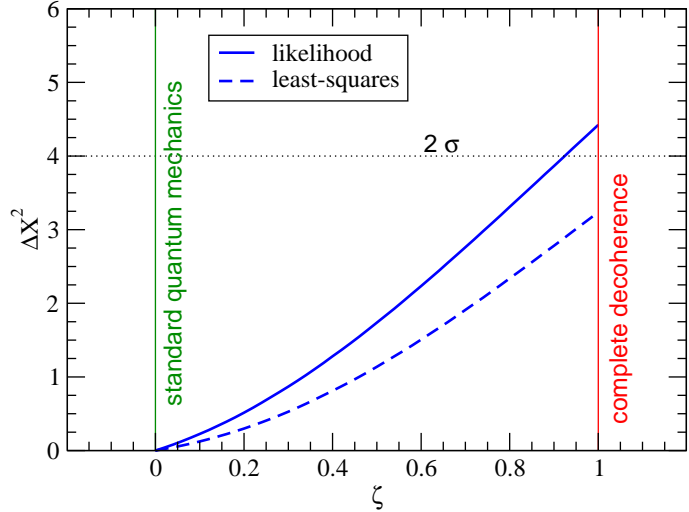}
\end{center}
  \caption{$\Delta X^2$ as a function of the decoherence parameter
  $\zeta$ for the likelihood and the least-squares method.}
  \label{fig:decoherence}
\end{figure}

Now the data is analysed as a function of the three parameters $\dm, \sq$ and
$\zeta$. The $\Delta X^2$ marginalised with respect to $\dm$ and $\sq$ is
shown in Fig.~\ref{fig:decoherence} for the likelihood and the least-squares
method. A clear indication in favour of oscillations is observed. Complete
decoherence is disfavoured with $\Delta X^2 = 4.4$ using the likelihood method
and $\Delta X^2 = 3.2$ by the least-squares method. This confirms that the
likelihood method is a more powerful tool to extract spectral information from
the data, in agreement with the results of Sec.~\ref{sec:KL_analyses}.
Assuming that $\Delta X^2(\zeta)$ is distributed as a $\chi^2$ with 1 degree
of freedom I conclude that the current KamLAND data provides a $\sim 2\sigma$
indication in favour of neutrino oscillations,\footnote{Note that here a {\it
relative} comparison of the fits for oscillations and decoherence is
performed; no statement about the absolute qualitiy of the fit is made. Hence,
these results are in agreement with Ref.~\cite{kamlandPRL}, where the observed
spectrum is found to be consistent with an oscillation signal at 93\%~CL but
with flat suppression at 53\%~CL.} implying quantum mechanical
interference over distances of the order of 200~km.

Depending on the true values of the oscillation parameters one may
expect that the statistical significance for oscillations in KamLAND
will strongly improve by future data. A simple rescaling of the
current data by a factor 5 leads to an exclusion of complete
decoherence at $3.7\sigma$ if the true value of $\dm$ turns out to be
$7\times 10^{-5}$~eV$^2$. On the other hand if $\dm = 1.5\times
10^{-4}$~eV$^2$ decoherence can be excluded only at $2.6\sigma$, since
for large $\dm$ the baselines in KamLAND are too long to be sensitive
to the oscillations.

\section{Conclusions}
\label{sec:conclusions}

In this letter two different analysis methods for the first data from the
KamLAND reactor neutrino experiment have been compared.  I found that an
event-by-event based likelihood method provides a more powerful tool to
extract information on two-neutrino oscillation parameters than a
least-squares method based on energy binned data. The likelihood method 
takes into account the precise energy information contained in each
single event and avoids the information loss due to binning.
Furthermore, exact frequentist confidence regions in the parameter space have
been calculated by means of Monte Carlo simulation according to the
Feldman-Cousins method~\cite{FC}. This method properly accounts for the
non-linearity of the oscillation parameters $\sq$ and $\dm$, and statistical
fluctuations in the data, which can be quite large due to the rather small
number of events in the current data sample. I have found a reasonable
agreement of the exact confidence regions with the ones obtained from the
$\chi^2$-cut approximation, especially in the vicinity of the best point at
$\dm \approx 7\times 10^{-5}$~eV$^2$. However, depending on the analysis
method (likelihood or least-squares) quantitative differences are visible,
especially for lower confidence levels and far from the best fit point.
Finally, although the current data is consistent with an energy independent
flux suppression, a $\sim 2\sigma$ indication in favour of oscillations can be
stated using the likelihood method, which is especially sensitive to the
spectral shape information. Put in other words, this implies a $\sim 2\sigma$
evidence for quantum mechanical interference over distances of the order of
200~km.

In summary, the results obtained in this work confirm that even for
the limited statistics of the current KamLAND data sample the
$\chi^2$-cut approximation to calculate confidence regions for the
oscillation parameters gives rather reliable results. One expects that in
future the agreement between approximate and exact confidence regions
will improve due to increase in statistics. Moreover, the differences
between likelihood and least-squares methods will become smaller.

{\bf Acknowledgements.} I thank M.~Maltoni and J.W.F.~Valle for
collaboration on the KamLAND analysis. Furthermore, I would like to
thank M.~Lindner, P.~Huber and T.~Lasserre for very useful
discussions. This work is supported by the ``Sonderforschungsbereich
375-95 f{\"u}r Astro-Teilchenphysik'' der Deutschen
Forschungsgemeinschaft.

\end{document}